\documentstyle[prd,preprint,aps,eqsecnum]{revtex}

\begin{document}
\draft
\title{Perturbative Approach at Finite Temperature and the $\phi^4$ model}

\author{G. Germ\'an}
\address{Instituto de F\'{\i}sica, Laboratorio de Cuernavaca,\\
Universidad Nacional Aut\'onoma de M\'exico,\\
Apartado Postal 48-3, 62251 Cuernavaca, Morelos, M\'exico\\}

\maketitle
\begin{abstract}
We suggest that the $\phi^4$ model is only a polynomial approximation to a more
fundamental theory. As a consequence the high temperature regime might not be
correctly described by this model. If this turns out to be true then several
results concerning e.g., critical temperatures, symmetry restoration at high
temperature and high temperature expansions should be reconsidered. We
illustrate our conjecture by using the Nambu-Goto string model. We compare a
two-loop calculation of the free energy or quark-antiquark static potential at
finite temperature with a previous exact calculation in the large-d limit and
show how the perturbative expansion fails to reproduce important features in
the neighborhood of the critical temperature. It becomes clear why this happens in
the Nambu-Goto model and we suggest that perhaps something similar occurs with
the $\phi^4$ model.
\end{abstract}

\newpage
\noindent

The purpose of this letter is to discuss the validity of the $\phi^4$ model in the
high temperature regime where symmetry restoration might occur. The $\phi^4$ model
is of great interest due to its wide range of applications, however, there are several
problems related with the model which might hint to the possibility that it is
only a polynomial approximation to a more fundamental theory. Although we have
no proof of this we would like to illustrate our conjecture by using the Nambu-Goto
string model as an example and discuss some issues related with the validity of the
perturbative expansion in the vecinity of the critical temperature. For this we
compare a two-loop calculation for a finite length Nambu-Goto string at arbitrary
temperature with a previous exact calculation of the free energy in the limit
of large-$d$ [1], $d$ being the number of dimensions of the embedding space,
where the string evolves.

The Nambu-Goto model is defined by the following action in Euclidean space
\begin{equation}
A=M^2\int d^{2}\xi\sqrt{g},
\end{equation}
where $M^2$ is the string tension and $g$ is the determinant of the metric
which is given in terms of the string coordinates by
\begin{equation}
g_{ij}=\partial_{i} x^{\mu}(\vec{\xi})\partial_{j} x^{\nu}(\vec{\xi})h_{\mu\nu},
\qquad i=0,1,
\end{equation}
being $h_{{\mu}{\nu}}$ the metric of the embedding Euclidean space where the string
evolves. For a $d$-dimensional space we have that $\mu$,$\nu$=$0,1,...,d-1$ and
$i,j=0,1$ for a string coordinate $x^\mu = x^\mu(\xi^0,\xi^1)=x^\mu(t,r)$.
We work in a Monge parametrization or "physical gauge"
\begin{equation}
x^{\mu}(\vec{\xi})=(t,r,u^a),
\end{equation}
where $u^a=u^a(t,r)$, $a=2,...,d-1$ are the $(d-2)$ transverse oscillating modes
of the string. The metric determinant can be written as [2]
\begin{equation}
g=detg_{ij}=1+\vec{u}^2_i+\frac{1}{2}\vec{u}^4_i-\frac{1}{2}(\vec{u}_i\cdot
\vec{u}_j)^2.
\end{equation}
For small field fluctuations and for the two-loop free energy we are interested
in, the action given by Eq.~(1) can be written as
\begin{equation}
A=M^2\int dtdr(1+\frac{1}{2}\vec{u}^2_i+\frac{1}{8}\vec{u}^4_i-
\frac{1}{4}(\vec{u}_i\cdot\vec{u}_j)^2).
\end{equation}
We now want to evaluate the functional integral
\begin{equation}
Z=\int Du^a e^{-A},
\end{equation}
for the quadratic part of the action. The interacting quartic terms are
evaluated in the usual way. The resulting expression for the effective action is
\begin{equation}
A_{eff}=\int drdt[M^2+\frac{d-2}{2}Trln(-\partial^2)]+<A_{int}>,
\end{equation}
where
\begin{equation}
<A_{int}>=\frac{(d-2)^2}{8}M^2\int drdt[(1-\frac{2}{d-2})<u_iu_i>^2-2<u_iu_j>^2].
\end{equation}
We now impose thermodynamic boundary conditions for a string with fixed ends at
finite temperature. The Green functions are
\begin{equation}
<u^a(t,r)u^b(t',r')>=\frac{2T\delta^{ab}}{M^2R}\sum^\infty_{n=1}\sum^\infty_{m=-\infty}
\frac{e^{i\omega_m(t-t')}}{k^2_n+\omega^2_m}sink_nrsink_nr',
\end{equation}
where $T$ is the temperature and $R$ the extrinsic length of the string. The
momenta and frequencies are given by
\begin{equation}
k_n=\frac{n\pi}{R}, \qquad \omega_m=2m{\pi}T \qquad n=1,2,... \qquad m=0,^+_-1,^+_-2,...,
\end{equation}
the trace is given by
\begin{equation}
Trln(-\partial^2)=\sum^\infty_{n=1}\sum^\infty_{m=-\infty}ln(k^2_n+\omega^2_m).
\end{equation}
A similar expression for the trace has been obtained before (see Eqn. (2.17) and
(2.20) of Ref. [1]). Also Eq.~(8) above can be calculated along the lines of
Ref. [2] but here the system is finite in both $R$ and $T$ directions. The resulting
expression for the free energy or static potential at finite temperature is
\begin{equation}
V(R,T)=V_0+V_1+V_2,
\end{equation}
where
\begin{mathletters}\label{13}
\begin{eqnarray}
V_0&=&M^2R\\
V_1&=&-\frac{(d-2)\pi}{24}\frac{1}{R}+
(d-2)T\sum^\infty_{n=1}ln(1-e^{\frac{-n\pi}{RT}})\\
V_2&=&-[\frac{(d-2)\pi}{24}]^2\frac{144}{2M^2R^3}[(\sum^\infty_{n=1}ncoth\frac{n\pi}{2RT})^2
+\frac{1}{d-2}\sum^\infty_{n=1}(ncoth\frac{n\pi}{2RT})^2
\nonumber\\
&&-(\frac{d-4}{d-2})\frac{RT}{\pi}\sum^\infty_{n=1}ncoth\frac{n\pi}{2RT}].
\end{eqnarray}
\end{mathletters}
An equivalent expression for $V$ has been calculated before in terms of Dedekind
functions by Dietz and Filk [3]. In that work analytic renormalization procedures
for functional integrals was the subject of interest. Here we would like to compare
with previous results in an exact large-$d$ evaluation of the effective potential
at finite temperature. The full potential, Eq.~(12), is now shown in Fig. 1 for several
values of the temperature in $d=4$ dimensions and with the zero temperature string tension
$M^2=M^2(T=0)$ normalized to unity. In the Nambu-Goto model, the string tension at
finite temperature for an infinitely long string was calculated some time ago by
Pisarski and Alvarez [4] with the result that there is a critical temperature
(the so-called deconfinement temperature) for which the string tension becomes vanishing, thus
signaling a transition to a deconfined phase. The value of this temperature is
\begin{equation}
T_{dec}=\sqrt{\frac{3}{(d-2)\pi}}|_{d=4}\sim 0.69.
\end{equation}
This result was obtained from and exact large-$d$ calculation for an infinitely long string
at arbitrary temperature. In an analogous calculation but for a finite length
string we showed [1] that for $T=T_{dec}$ the potential becomes flat for large $R$ thus losing
the "confinement" property or linear behavior of the string (see inset of Fig. 2a of Ref. [1]).
This, as we argued before, is a perfectly consistent result with the identification of
$T_{dec}$ as a deconfinement temperature. In our two-loop calculation, however,
this result does not show up. We investigated numerically the curve of Fig. 1 with
$T=0.69$ never becoming flat for large $R$. A simple expansion of Eq.~(12) for
large $R$ and the use of $Z$-function regularization to evaluate the sums
$\sum^\infty_{n=1}=\zeta(0)=-\frac{1}{2}$ and $\sum^\infty_{n=1}\ln n=-\zeta'(0)=\frac{1}{2}\ln2\pi$
shows that in fact the potential behaves like
\begin{equation}
V(R,T)\sim M^2R-\frac{(d-2)\pi}{6}RT^2+\frac{(d-2)}{2}T\ln2RT,
\end{equation}
thus never showing the constant potential obtained in the exact calculation. For
temperatures bigger than the the critical value Fig. 2b of Ref. [1] shows
that the potential exists although for a string up to certain length. In [1] this
was understood as follows: for large $R$
\begin{equation}
V(R\rightarrow \infty,T)\sim RM^2(R\rightarrow \infty,T)=RM^2(T)=R\sqrt{1-\frac{T^2}{T^2_{dec}}}.
\end{equation}
For $T=T_{dec}$, $M^2(T)=0$ and in the competition of limits with $R\rightarrow \infty$
$V(R\rightarrow \infty,T\rightarrow T_{dec})$ the potential results in a constant with an
approximated value of $6.2$. For $T\geq T_{dec}$ the potential starts "seeing"
the linear behavior in $R$ with a slope becoming increasingly close to
$\sqrt{1-\frac{T^2}{T^2_{dec}}}$ thus appearing a negative radicand which "stops" the
potential for a given $R$. In general the expression for the exact potential at
finite temperature should be a very complicated function of $R$ and $T$ but for
large $R$ a simpler expression of the form (16) should emerge. Thus for
$T>T_{dec}$ we can no longer have strings of arbitrary length. In our
two-loop result we see in Fig. 1 (curves with $T=0.75$ and $0.8$) that this interesting
feature is not present as we can also see from the large-$R$ expansion of Eq.~(15).
In Fig. 2 we plot the ratio of the two-loop correction over the one-loop contribution
for the same values of the temperature shown in Fig. 1. We see that for certain values
of $R$ this ratio becomes bigger than one thus invalidating the loop calculation.
We also see that the values of $R$ for which the potential stops in Fig. 2b of
Ref. [1] lie in the region for which $V_2/V_1>1$, this ratio becoming again well
behaved for larger values of $R$ which, however, is an artefact of the perturbative
expansion; in the exact treatment of the problem this region does not exist,
the string having probably decayed. Note that something similar happens in the
$\phi^4$ model close to the critical temperature $T_c$ where $m_\phi(T)\rightarrow 0$.
Here the problem is usually dealt with by including and infinite set of daisy and
superdaisy diagrams. It is usually said that for $T>T_c$ higher-order corrections
are again small and the results reliable. However, it has been argued recently that the
$\phi^4$ model has no particle interpretation for $m^2_\phi>0$, for large
temperatures the $\phi$-particle probably disappears [5,6]. In the Nambu-Goto
model we have seen that for $T>T_{dec}$ the ratio $V_2/V_1$ is also well behaved
to the right of the peak, however because we are able to compare with the exact result
we do not trust this region; as we said before it is only an artefact. Thus the
perturbative result is jumping the singularity of the exact potential not showing
any of the interesting behavior in the neighborhood of the critical temperature.
It is easy to see why this happens: instead of beginning with Eq.~(1) above we could
have started with a model given by Eq.~(5) without asking about its origin. Clearly
everything would follow exactly the same and we would get Figs. 1 and 2 with the peculiar
behavior noted before and we could agree that perturbation theory for $T>T_{dec}$
is playing a trick on us. Using the model given by Eq.~(5) we could believe that, because
$V_2/V_1>1$ only in a small region of $R$ values, everywhere else perturbation theory
is fine even for $T>T_{dec}$. As it is, however, we can easily trace back the jumping
of the singularity by the perturbative approach to the fact that from the very
beginning we are expanding $\sqrt g$ in Eq.~(1) to get Eq.~(5), i.e., in a way Fig. 2
is telling us that we are working with an approximated model given by Eq.~(5) instead
of the full one Eq.~(1). Having the model Eq.~(1) and the exact treatment of the
problem [1] we can see that in fact, for $T>T_{dec}$, all the region to the right
of the peak $V_2/V_1>1$ where the ratio $V_2/V_1$ becomes again well behaved
(see Fig. 2) is no longer reliable. Actually this phase is not even described by the Nambu-Goto
model.

Now let us suppose that we are working with the $\phi^4$ model and that we find
something similar to Fig. 2 but now with $V_2(\phi,T)/V_1(\phi,T)$ versus $\phi$
for various values of $T$. Our suggestion is that a similar behavior to the one
described in this letter might be hinting towards a more fundamental origin of
our field theory, something playing the role of our Eq.~(1) above, and that the "failure"
of the perturbative approach is in fact telling us that we are working with an
approximated expanded model of a more fundamental theory. As a consequence the
high temperature regime might not be correctly described by the $\phi^4$ model.
If this turns out to be correct (and it should be investigated) then several
results concerning e.g., critical temperatures, symmetry restoration at high temperature
and high temperature expansions should be reconsidered.

\newpage
\noindent
{\bf Figure Captions}

\bigskip
\noindent
Fig.1\\
The two-loop effective potential for the Nambu-Goto string model is shown for
various values of the temperature. If we compare with Fig. 2b of Ref. [1] we
see that the perturbative calculation does not show the interesting behavior
in the neighborhood of the critical temperature $T_{dec}$. The perturbative
expansion "jumps" the singularity present in the exact result at $T=T_{dec}$.

\bigskip
\noindent
Fig.2\\
The two to one loop corrections ratio is here shown as a function of the string
length for various values of the temperature. The perturbative expansion is no
longer reliable when the temperature is in the vecinity of the critical value.
The value of $R$ for which the potential ceases to exist in the exact result
(Fig. 2b of Ref. [1]) lies in the region where $V_2/V_1>1$ above. In the
perturbative approach the potential does not notice that it should not exist
for some $R$ when $T>T_{dec}$, where the string has probably decayed.

\newpage
{\bf References }

\medskip
\begin{itemize}
\item[1)] A. Antill\'on and G. Germ\'an, Phys. Rev. D{\bf 47}, 4567(1993).

\item[2)] G. Germ\'an and H. Kleinert, Phys. Rev. D{\bf 40}, 1108(1989).

\item[3)] K. Dietz and T. Filk, Phys. Rev. D{\bf 27}, 2944(1983).

\item[4)] R.D. Pisarski and O. Alvarez, Phys. Rev. D{\bf 26}, 3735(1982).

\item[5)] H.A. Al-Kuwari, Phys. Lett. {\bf 375B}, 217(1996).

\item[6)] B.A. Campbell, J. Ellis and K.A. Olive, Phys. Lett {\bf 235B}, 325 (1990).

\end{itemize}

\end{document}